\documentclass[aps,prb,superscriptaddress, amsmath,amssymb,twocolumn,a4paper,showpacs]{revtex4}

\usepackage[pdftex,breaklinks=true,colorlinks=true]{hyperref}
\usepackage[pdftex]{graphicx}

\usepackage{braket}
\begin{document}

\title{Flux quench in a system of  interacting spinless fermions in one dimension}

\author{Yuya O. Nakagawa}
\email{y-nakagawa@issp.u-tokyo.ac.jp}
\affiliation{Institute for Solid State Physics, the University of Tokyo, 
5-1-5 Kashiwanoha, Kashiwa, Chiba Japan 277-8581 }

\author{Gr\'egoire Misguich}
\affiliation{Institut de Physique Th\'eorique, Universit\'e Paris Saclay, CEA, CNRS, F-91191 Gif-sur-Yvette, France}

\author{Masaki Oshikawa}
\affiliation{Institute for Solid State Physics, the University of Tokyo, 
5-1-5 Kashiwanoha, Kashiwa, Chiba Japan 277-8581 }

\date{\today}

\begin{abstract}
We study a quantum quench
in a one-dimensional spinless fermion model (equivalent to the XXZ spin chain),
where a magnetic flux is suddenly switched off. This quench is equivalent to imposing a pulse of
 electric field  and therefore generates an initial particle current.
This  current is not a conserved quantity in presence of a lattice and interactions and
we investigate numerically its time-evolution after the quench, using the infinite time-evolving block decimation method.
For repulsive interactions or large initial flux, we find oscillations
that are governed by excitations deep inside the Fermi sea.
At long times we observe that the current remains non-vanishing in the gapless cases, whereas it decays to zero in the gapped cases.
Although the linear response theory (valid for a weak flux) predicts the same long-time limit of the current for repulsive and attractive interactions (relation with the zero-temperature Drude weight), 
larger nonlinearities are observed in the case of repulsive interactions compared with that of the attractive case.
\end{abstract}


\maketitle


\section{\label{Introduction} Introduction }
The nonequilibrium dynamics of isolated quantum systems has become a major subject of
study in condensed matter physics~\cite{Polkovnikov_RMP}.
Thanks to substantial developments on the experimental side,
it is now possible to compare theoretical predictions and experiments with a high accuracy and controllability,
in particular in the field of cold atoms. Quantum quench, a sudden change of 
some parameter(s) in a quantum system, is one of the simplest protocols to
drive systems out of equilibrium.
Typically, an initial state is prepared as the ground state of some pre-quench Hamiltonian
and some external parameter is then abruptly changed at $t=0$. This leads, for $t>0$,
to a unitary evolution with a different Hamiltonian and to some non-trivial dynamics.

Quantum quenches in one-dimensional (1d) systems have already been intensively studied
for several reasons.
First, the effect of interactions and quantum fluctuations are particularly 
important in 1d. Second, several powerful analytical and numerical
methods, such as Bethe Ansatz, bosonization, Time-Evolving Block
Decimation (TEBD), and time-dependent Density Matrix Renormalization
Group (t-DMRG),
are available for these systems;
these methods allow us to make predictions concerning
the dynamics of these quantum many-body systems.
In the present study we consider a simple quench for interacting spinless fermions in 1d,
where an Aharonov-Bohm   flux is suddenly switched off.
This is equivalent to an application of an instantaneous pulse of
electric field, which generates an initial particle current.
This quench has several appealing properties. First, the non-trivial dynamics comes specifically from lattice and interactions effects.
Indeed, the current is an exactly conserved quantity for free fermions on a lattice, as well as for any model with translation
symmetry in the continuum (due to Galilean invariance)~\footnote{In these two cases the sudden change of flux creates an excited {\em eigenstate} (a boosted Fermi sea for free fermions) and
no dynamics is generated (see also Appendix~\ref{sec:twist}).
In presence of a lattice {\em and} interactions, Umklapp processes can change the particle current
while conserving the total lattice momentum.}.
So, it is the combination of interactions and lattice effects that is responsible for
the nontrivial dynamics.
Second, changing the flux by an integer number of flux quanta on a periodic chain amounts
to a unitary transformation of the Hamiltonian and therefore leaves
the energy spectrum unchanged. In other words, for an integer number of flux quanta, the energy spectrum
is the same for the pre-quench and post-quench Hamiltonians. In that case the dynamics solely comes from a change in the eigenstates, not from their energies.
Third, this quench allows to make contact with some transport properties of the system.
When the number of flux quanta per the length of the system is small,
the electric field pulse is weak. In this limit,
the dynamics may be described by using the linear response theory;
long-time limit of the current should then be
directly proportional to the zero-temperature
Drude weight of the model \cite{oshikawa_insulator_2003,Mierzejeski2014}.

In this study we use the infinite time-evolving block decimation (iTEBD) \cite{Vidal2007} method to monitor the evolution of the wave function.
We focus on the particle current and analyze its dynamics, including its long-time limit.
As an important result, we observe some current oscillations at intermediate times.
In addition, these current oscillations are found to be carried by excitations located deep inside the Fermi sea.
Finally, we find that the long-time limit of the current depends in a nonlinear way on the initial flux. 
These nonlinearities appear to be particularly strong in the case of repulsive interactions between the particles.
A theoretical understanding of these observations -- presently lacking -- would require us to go beyond an effective low-energy description, such as bosonization.
Several other quantities, like the growth of the entanglement entropy, are also computed.

The remainder of the paper is organized as follows.
In Sec.~\ref{Model}, we introduce the model and the flux quench problem. 
In Sec.~\ref{Results},
we  review the numerical method (iTEBD) and present our numerical results for the dynamics after the quench.
In Sec.~\ref{Conclusion}, we summarize our results and state conclusions.
Technical details on numerical calculations are presented in the Appendix.

\section{\label{Model} Flux quench}

We consider one of the simplest interacting spinless fermion systems in
one dimension (assuming a periodic boundary condition):
\begin{equation} \label{DefOfModel} 
 H(t) = - \frac{1}{2} \sum_{i=0}^{N-1} \left( 
e^{-i\theta(t)} c^\dagger_i c_{i+1}  + \text{h.c.}  \right)
- \Delta \sum_{i=0}^{N-1}  \tilde{n}_i \tilde{n}_{i+1},
\end{equation} 
where $\tilde{n}_i = c^\dagger_i c_{i}-1/2$,
and $N$ is a total number of sites.
As is well known, a Jordan-Wigner transformation maps this
model to  a spin-$1/2$ XXZ chain~\cite{GiamarchiBook}.
We focus here on the zero chemical potential 
case, which corresponds to zero external magnetic field
in the spin language.
The phase factor $\theta(t)$ in the hopping terms is
the vector potential, representing
an Aharonov-Bohm flux $\Phi(t)=N\theta(t)$ piercing the ring.
In the spin language, it introduces a twist in the $xy$-plane.
In the following, we call $\theta(t)$ flux strength.

In what follows we take the thermodynamic limit $N\to\infty$ while keeping
the flux strength $\theta(t)$ constant.
The model is integrable for any value of $\Delta$, and the phase diagram
has been thoroughly studied~\cite{GiamarchiBook}.
For $\Delta > 1 \: (\Delta =1)$, the system is gapped (gapless), and there are two degenerate
ground states which are
exactly given by the completely empty state and the completely filled
state. They correspond to ferromagnetic states in the spin language.
In these cases, the ground states are completely insensitive to the flux
and there will be no dynamics as well.
For $-1 \leq \Delta < 1$, the system is gapless and its low-energy
universal behaviors are described by bosonization as Tomonaga-Luttinger
liquid.
For $\Delta < -1$, the system is again gapped, and there are two
degenerate ground states corresponding to an antiferromagnetic
long-range order in the spin language.
In contrast to the ferromagnetic case, however, the ground states
are still nontrivial and sensitive to the flux.
Therefore, in this paper, we consider the regime $\Delta <1$,
where there are nontrivial effects of the flux.

{\it Flux quench} -- The problem we study here is
a quantum quench where the flux $\theta$ is varied from $\theta(t<0) =  \theta_0$ to $\theta(t\geq 0) =0$.
This sudden change of magnetic flux is equivalent to imposing an instantaneous pulse of electric field 
$E(t) = -\partial_t \theta(t) = \theta_0\delta(t) $ to the  fermions, and it induces some particle current
at the initial time.
In the present setup, the current is always uniform throughout the
system and thus we can define the current operator by the
average of local currents as~\footnote{In presence of
a non-zero vector potential $\theta(t)$, the expression of the particle current
is modified to $\hat{J} = \frac{1}{2iN}\sum_i  \left( e^{-i\theta(t)} c^\dagger_i c_{i+1} - e^{i\theta(t)} c^\dagger_{i+1}c_i\right)$
in order to be gauge-invariant.}
\begin{equation}
 \hat{J} = \frac{1}{2iN}\sum_i  \left( c^\dagger_i c_{i+1} - c^\dagger_{i+1}c_i\right)
 = \frac{1}{N}\sum_q  \sin(q) \tilde{c}^\dagger_q \tilde{c}_q,
\end{equation}
where 
$  \tilde{c}_q := \frac{1}{\sqrt{N}} \sum_{r=0}^{N-1} c_r e^{-iqr}$
is the annihilation operator in momentum space.
The current is not a constant of motion in presence of
interaction $\Delta \neq 0$. The latter, combined with the presence of
lattice, causes Umklapp scattering as
\begin{equation}
 [H_0, \hat{J}] =  -\frac{\Delta}{2iN} \sum_i
   \left( c_i^\dagger c_{i+1} + c_{i+1}^\dagger c_i   \right)
   \left( n_{i-1} - n_{i+2}  \right), \label{dJdt}  
   \end{equation}
where $H_0$ is the Hamiltonian without flux, $H_0 = H(t \geq 0)$.
As mentioned in the introduction, the interactions are  essential to produce a nontrivial dynamics.
In this study we focus on the expectation value
of the current, $J(t) := \langle \psi(t) | \hat{J} | \psi(t)\rangle$. In particular we analyze (i) the time-evolution towards stationary states and (ii) the long-time limit of the current.
We note that Mierzejeski {\it et al.} \cite{Mierzejeski2014} recently utilized this flux quench
to illustrate the breakdown of the generalized Gibbs ensemble \cite{Rigol_GGE2007}.
Also, the Loschmidt echo associated to this quench
was considered using the Bethe ansatz in Ref.~\onlinecite{de_luca_quenching_2014}, and
the flux quench for bosons was studied in Refs.~\onlinecite{Danshita2010,Danshita2011}.

When $\theta_0$ is small,
we expect that the linear response (LR) theory can be applied
to obtain $J(t)$ as a response to the weak electric pulse as
\begin{equation}
  J(t) = \frac{1}{2\pi} \int_{-\infty}^{\infty} d\omega F(\omega) \sigma(\omega)e^{-i\omega t}+ O(\theta_0^2),
  \label{eq:LR}
\end{equation}
 where $F(\omega) = \theta_0$ is the Fourier transform
of the imposed electric field $E(t)= \theta_0\delta(t)$,
and $\sigma(\omega)$ is the conductivity%
\begin{equation}
 \sigma(\omega) = \frac{N}{\omega}
 \int_0^{\infty} dt \, e^{i\omega t}  \langle [\hat{J}(t), \hat{J}] \rangle_{\mathrm{GS},0}.
\end{equation}
Here $\langle \ldots \rangle_{\mathrm{GS},0}$ denotes the expectation in
the ground state of $H_0$.
The conductivity has a zero-frequency component, called the Drude weight, as well as a regular part:
\begin{equation}
 \sigma(\omega) = 2\pi D\delta(\omega) + \sigma_\mathrm{reg} (\omega).
\end{equation}
In the case of flux quench, Eq.~\eqref{eq:LR} gives
\begin{equation}
    J(t) = \frac{\theta_0}{2\pi} \int_{-\infty}^{\infty} d\omega \sigma(\omega)e^{-i\omega t}+ O(\theta_0^2)
\end{equation}
and, as noted in Ref.~\onlinecite{Mierzejeski2014}, the long-time limit of the current is proportional to the Drude weight $D$
\begin{equation}
 J(t=\infty) = D\theta_0, \label{DrudePrediction}
\end{equation}
while the finite-time dynamics  is governed by the regular
part of the conductivity $\sigma_\mathrm{reg} (\omega)$.
We will compare our numerical data on the flux quench with these LR predictions later.

Before analyzing this problem in detail,
we mention some possible experimental realizations.
The flux quench can be viewed as a sudden momentum shift for the particles.
It is therefore equivalent to a situation where a moving lattice stops abruptly at $t=0$ (the lattice velocity
provides the initial momentum shift).
This situation was experimentally realized \cite{BosonExperiment}
with bosons trapped in an optical lattice. We may therefore expect that a similar setup could
be realized with fermions~(\ref{DefOfModel}).
Besides, a quantum quench using an artificial gauge field in optical lattices was also
proposed \cite{Peotta2014}. This is a direct realization of the flux quench studied here, although
bosons were considered.

\section{\label{Results} Numerical Results}
In this section we present numerical results for the dynamics of the current.
We employ the iTEBD
method~\cite{Vidal2007}, which enables  to study the system in
thermodynamic limit $N \to \infty$.
The iTEBD is a numerical scheme based on the Matrix-Product State (MPS)
representation of quantum many-body states in 1d.
The MPS can naturally describe a translationally invariant state
of an infinitely long 1d system.
In the present problem, the initial state is translation invariant,
and this symmetry is preserved also at $t>0$ by
the post-quench Hamiltonian $H_0$.
Thus, strictly speaking, there is no finite-size effect in our calculation.
On the other hand, an exact description of a given quantum state
by an MPS generally requires a matrix of infinite dimensions, but we need
to approximate it by a finite-dimensional matrix in a practical
calculation. The dimension of the matrix is called bond dimension,
and the use of a finite bond dimension is a possible source of the error
in the calculation.
For a ground state of a gapped 1d system, an MPS with a (sufficiently large) finite bond
dimension is known to provide an almost exact description of the wave function~\cite{Hastings_arealaw_2007}.
However, for the ground state of a gapless system with an infinite correlation length,
the finite bond dimension of an MPS approximation is known to introduce
an effective finite correlation length~\cite{tagliacozzo_scaling_2008}.
Nevertheless, the MPS description
(and thus iTEBD algorithm)  with a finite bond dimension
can provide an accurate result on quantum dynamics up to a certain
time~\cite{Pollmann2013}, in particular concerning local observables.
Therefore we apply the iTEBD algorithm to the flux quench
problem, to obtain the evolution of the current for a certain
period of time after the quench.

In practice, first, the ground state of the model~(\ref{DefOfModel}) with
flux $\theta_0$ is obtained by simulating an imaginary time-evolution
with the iTEBD algorithm. We then compute the real time-evolution using
the Hamiltonian without flux, still with iTEBD.  We carefully
check the numerical errors by varying the time steps during the
time-evolution, as well as the bond dimension $\chi$. The
results shown in the present paper were obtained by using bond dimensions
between $\chi=500$ and $\chi=1200$.  We calculate the dynamics for various
initial flux strengths $\theta_0$ ranging from $\pi/30$ to $\pi/2$, and for
interaction $\Delta$ from $-2.0$ to $0.8$.  More details on the
numerical calculations are given in Appendix \ref{sec:app_num}.

\begin{figure*}
   \includegraphics[width=8cm]{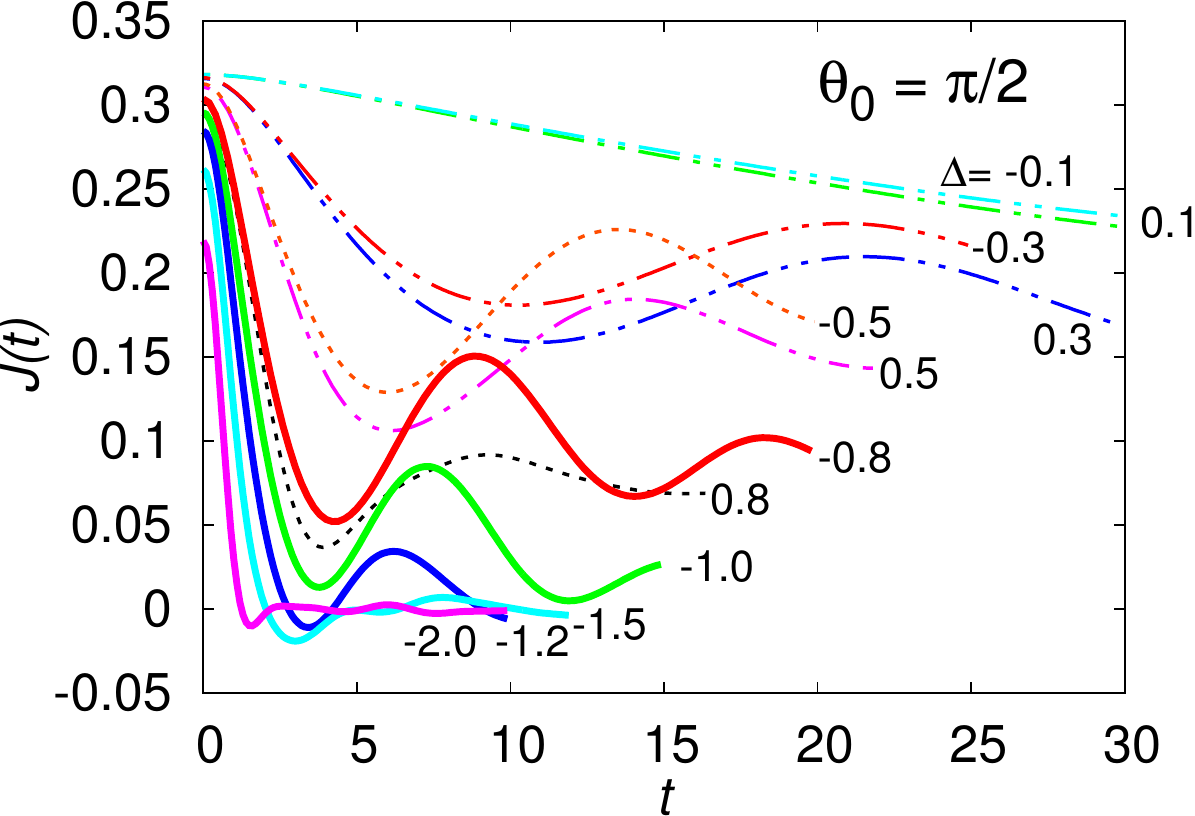}
   \includegraphics[width=8cm]{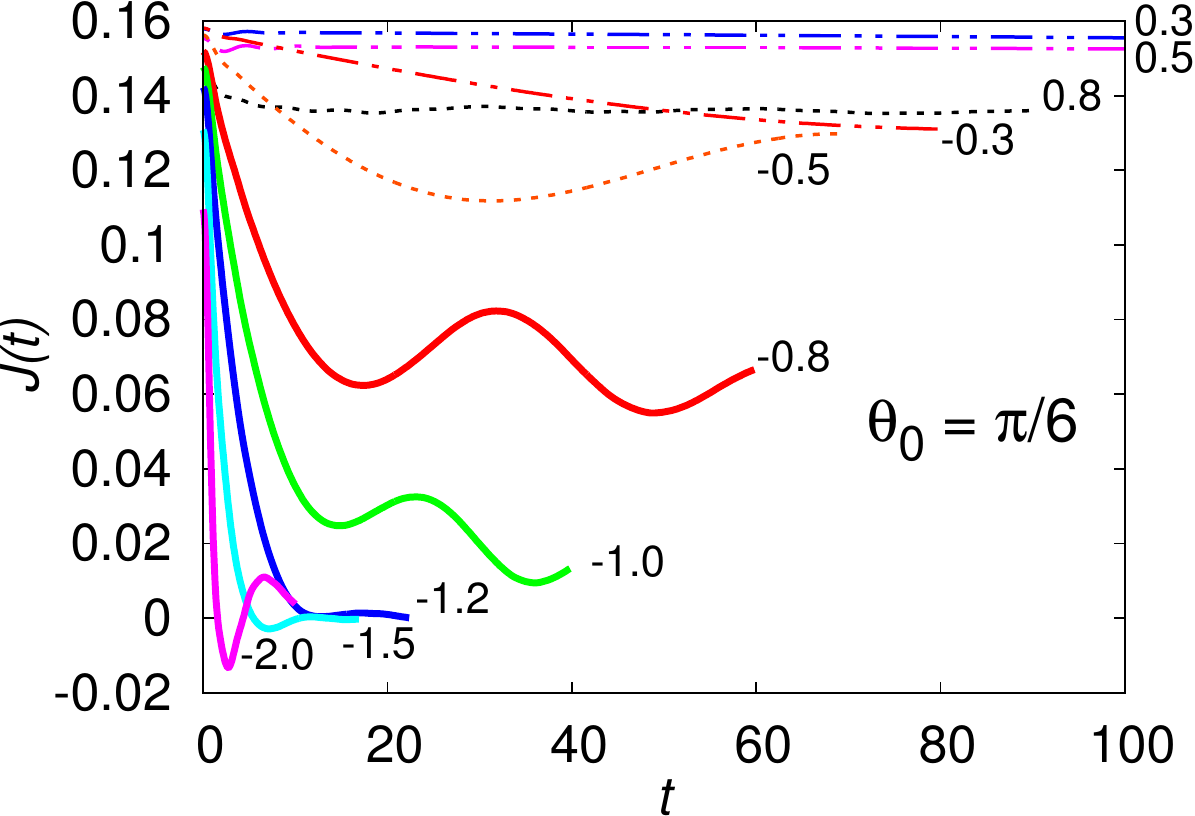}
   \includegraphics[width=8cm]{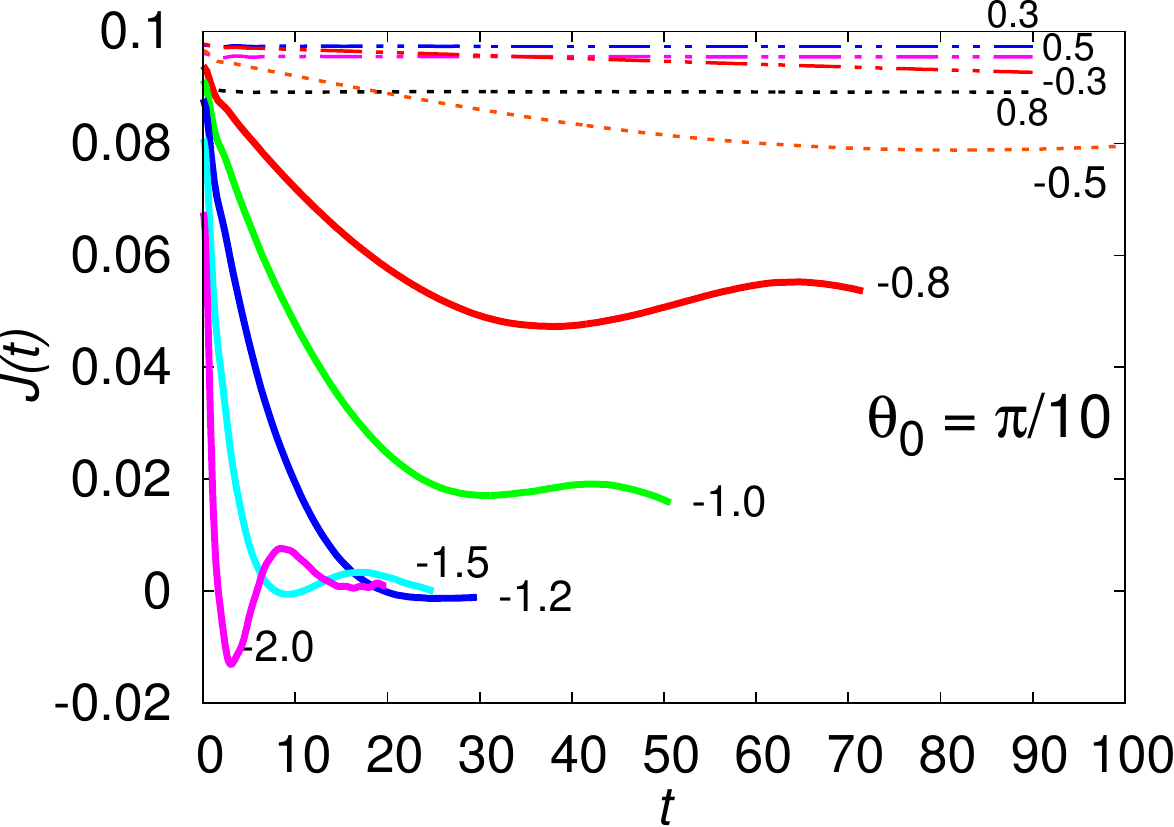} 
   \includegraphics[width=8cm]{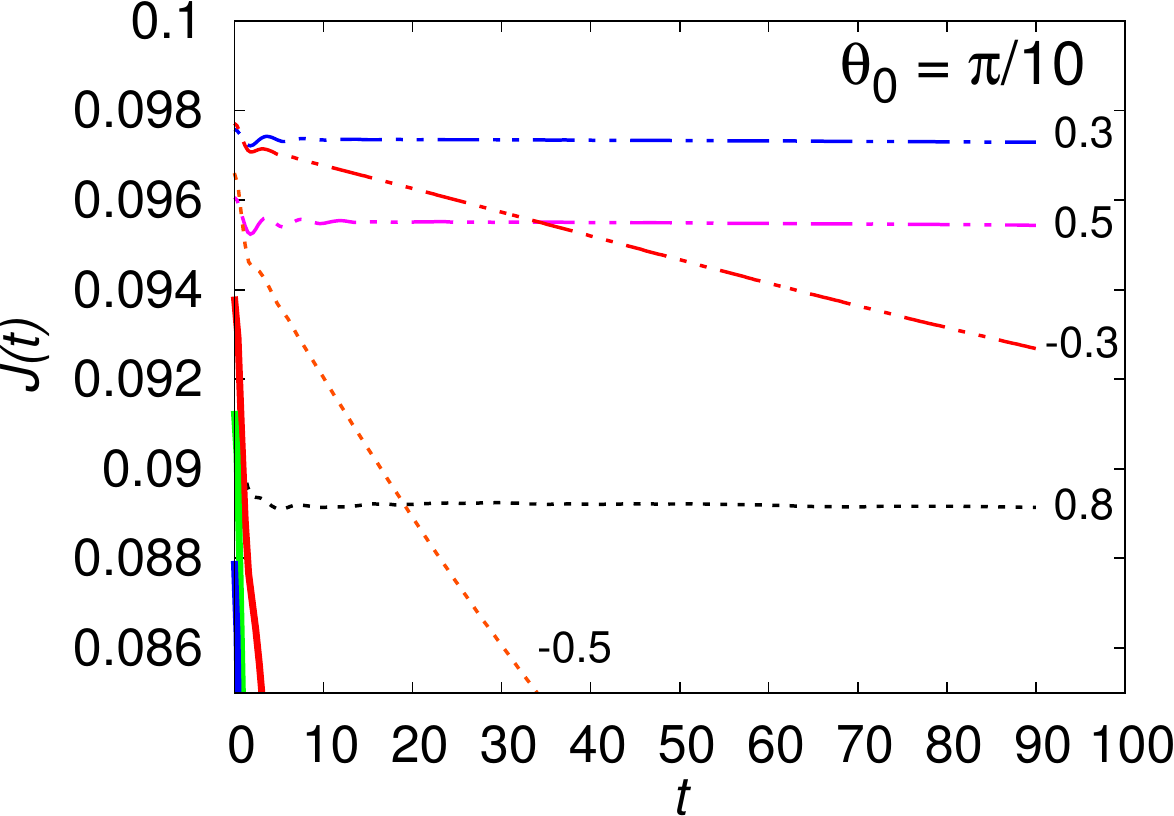}
  \caption{\label{Result1}
  Dynamics of the current after the quench for  $\theta_0=\pi/2$ (upper left),
  $\theta_0=\pi/6$ (upper right), and $\theta_0=\pi/10$ (bottom left and right).}
\end{figure*}

The evolution of the expectation value of the current, $J(t)$, is shown
in Fig.~\ref{Result1}.  We summarize the observed dynamics as follows.
For large initial flux $\theta_0$ ($\theta_0 \gtrsim \pi/3$), the
current shows some decay and oscillations for all values of interaction
$\Delta$ (note however that the oscillation period for $\Delta = \pm
0.1$ is too long to be measured accurately).  For smaller initial
fluxes, $\theta_0 = \pi/6$ to $\pi/30$, we observe a qualitatively
different dynamics, depending on the sign of $\Delta$.  For attractive
interactions ($\Delta >0$), the oscillations (if any) are too slow to be
visible within the simulation time, and the relative decay of the
current is small. In that regime $J(t)$ quickly reaches a stationary
value (with the possibility of some short time-scale and small amplitude
oscillations, as visible in the inset of the right panel of
Fig.~\ref{TimeEvol_MomD}).
For repulsive interactions ($\Delta <0$), some oscillations are visible, although for small initial flux $\theta_0$
and small $|\Delta|$ their period can exceed the simulation time.
Besides, the decay of the current is larger and the associated relaxation time scale 
is longer than in the attractive case.

\begin{figure}
  \includegraphics[width=8cm]{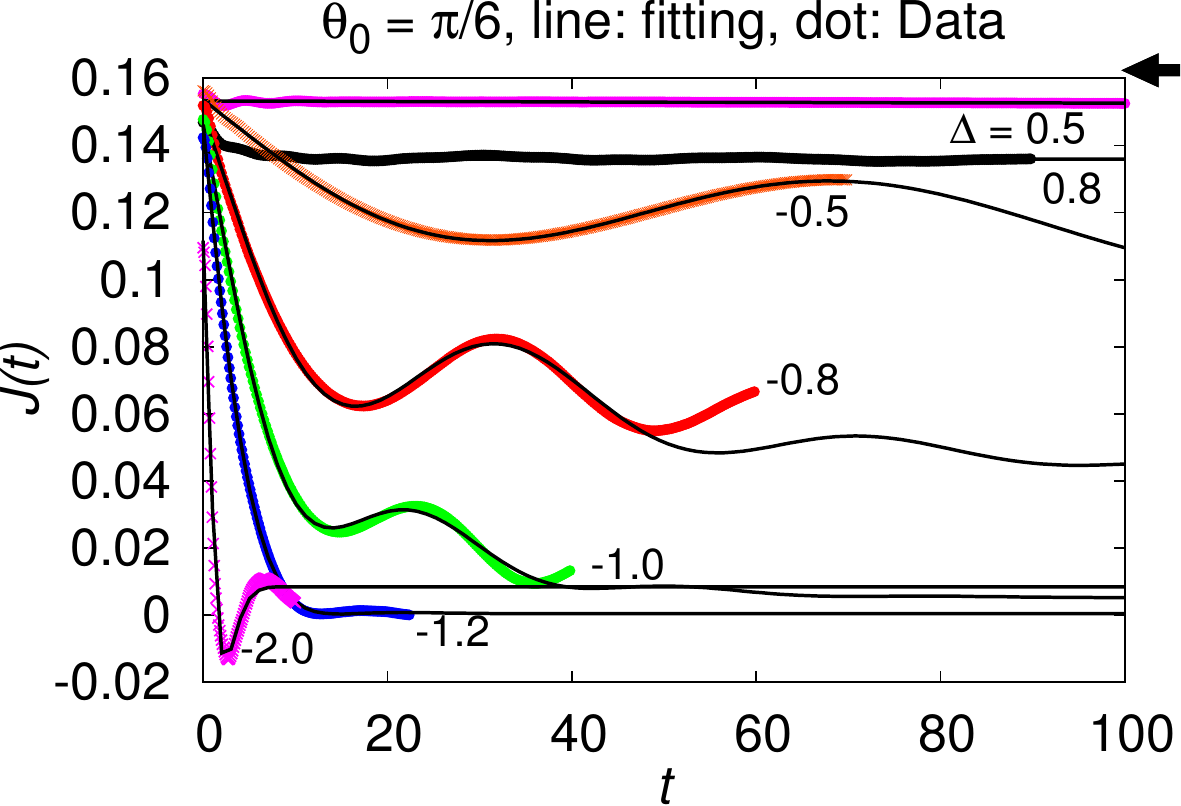}
  \caption{ \label{ExampleOfFit} Fitting of numerical data for initial flux $\theta_0 = \pi/6$.
  Roughly up to first oscillation, the empirical fitting works well.
  The black arrow indicates the LR prediction of the long-time limit of the current for $\Delta = \pm 0.5$
  (Eqn.~\eqref{DrudePrediction}). }
\end{figure}

In order to quantify the various scales associated to the current dynamics,
we use two simple fitting functions,
\begin{equation}
  \begin{cases}
  f(t) = c + \left( A + B \cos(\omega t + \phi) \right) e^{-t/ \tau}  \\
  g(t) = c + A e^{-t/ \tau} 
  \end{cases},
\end{equation}
where $c,A,B,\omega,\phi,\tau$ are fitting parameters.
We use $f(t)$ when some oscillations are visible within the simulation time scale ($t < 100$),  and 
$g(t)$ otherwise.
Some examples of fits are shown in Fig. \ref{ExampleOfFit}.
Among the fitting parameters, we focus on $c, \omega$ and $\tau$,
which correspond respectively to the long-time limit of the current,
the frequency of the oscillations, and the relaxation time.

\subsection{Oscillations}

\begin{figure}
   \includegraphics[width=8cm]{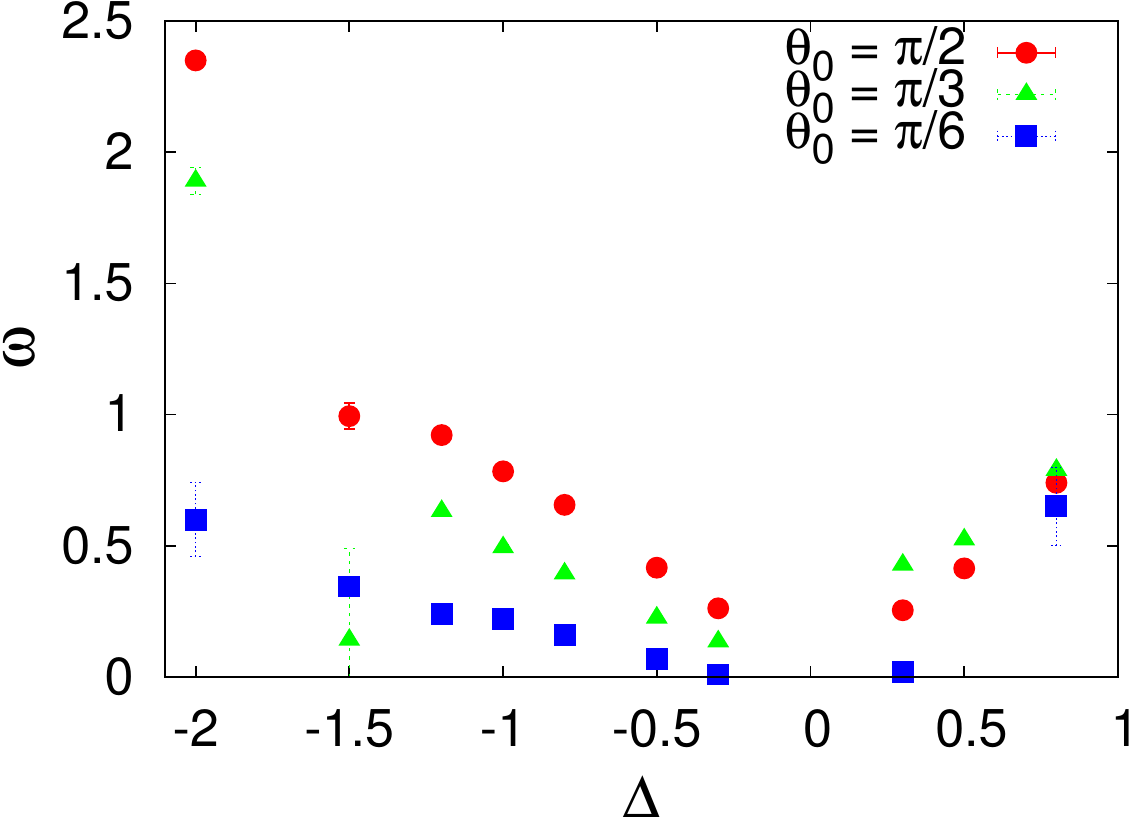}
   \caption{ \label{Omega}  Frequency $\omega$ from numerical fitting. Error bar is estimated
   by error of fitting.} 
\end{figure}

As shown in Fig.~\ref{Omega}, the 
oscillation frequency $\omega$ extracted from the fits is approximately linear in $|\Delta|$. 
Note however, the associated slope appears to depend on the value of the initial flux.
The relation $\omega \propto |\Delta|$ seems to hold in
gapless phase ($|\Delta| \leq 1$).
But it may also be valid beyond that regime, since $\omega$  also appears to be approximately
linear $|\Delta|$ in the regime $-1.5 \lesssim \Delta \leq 1$, although this relation clearly breaks down for
$\Delta \lesssim -1.5$.

We comment on the relation between our numerical results and the LR theory.
As described in the previous section, the LR theory ($\theta_0 \ll 1$)
relates the real-time dynamics of the current to the Fourier transform of the regular part of
the conductivity $\sigma_\mathrm{reg} (\omega)$.
Even though the applicability of the LR theory is not obvious when $\theta_0$ is of the order of unity
(as for $\theta_0 = \pi/6, \pi/3$ and $\pi/2$), the frequency $\omega$ of the observed oscillations may originate from a peak
in $\sigma_\mathrm{reg} (\omega)$.
To our knowledge such a structure in $\sigma_\mathrm{reg} (\omega)$ has not been explicitly discussed in the literature for the XXZ model,
but similar results have been reported in studies on the finite-temperature Drude weight
of this model \cite{Herbrych2011}.
Also, it is worthwhile to point out that the current-current  correlation function
$\langle \hat{J}(t) \hat{J} \rangle$
shows a similar oscillatory behavior \cite{OscillatoryCorrelation}
(basically the integral of this correlation function gives the time-dependence of $J(t)$).

\subsubsection{Dynamics of the momentum distribution}
\begin{figure*}
   \includegraphics[width=8cm]{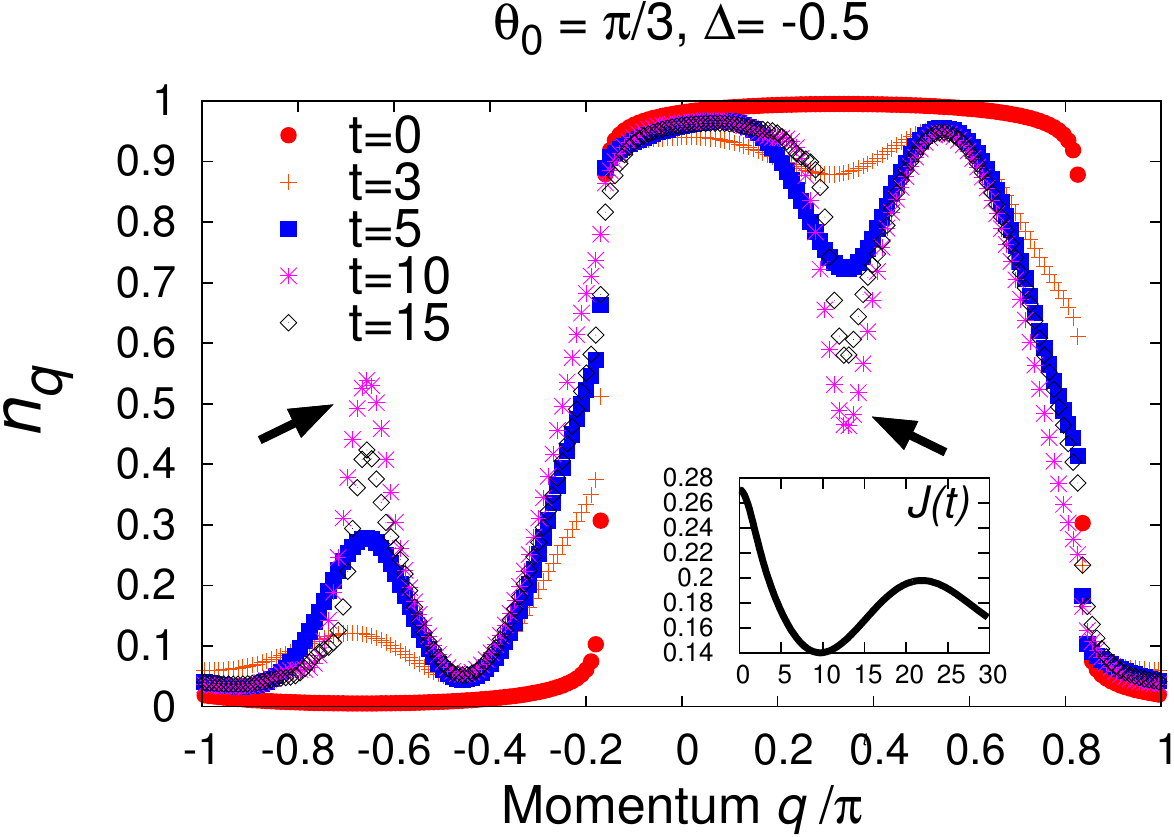}
    \includegraphics[width=8cm]{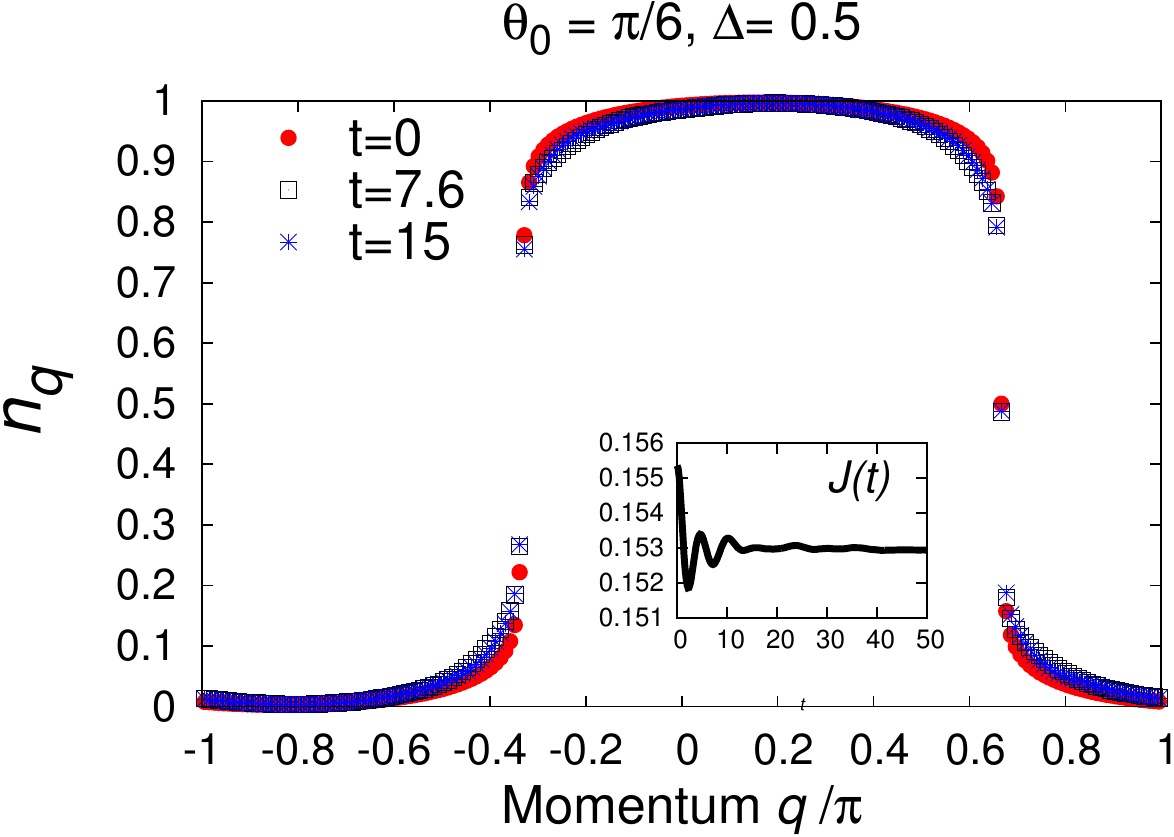}
  \caption{  \label{TimeEvol_MomD}
  Time evolution of the momentum distribution $n_q$ for (left) $\theta_0= \pi/3,\,\Delta=-0.5$ and
  (right) $\theta_0= \pi/6,\,\Delta=0.5$.
  An oscillating dip/peak structure is visible in the left panel (marked by arrows), whereas such
  structure is absent in the right case. See also the Supplemental Material~\cite{SuppMat}.}
\end{figure*}

To investigate the nature of the current oscillations, we calculate the momentum distribution
of the particles, $n_q = \langle \tilde{c}^\dagger_q \tilde{c}_q \rangle$.
Since we have  $\langle J(t) \rangle = \frac{1}{N}\sum_q \sin(q)n_q(t)$,
a population imbalance between $q>0$ and $q<0$ results in non-zero current.
Figure.~\ref{TimeEvol_MomD} shows the dynamics of the momentum distribution
for two cases~\cite{SuppMat}:~$\theta_0 = \pi/3, \,\Delta = -0.5$ and $\theta_0 = \pi/6, \,\Delta = 0.5$.
At $t=0$ the momentum distribution is
that of the ground state in presence of a magnetic flux. It corresponds to the ground state in zero flux, but shifted by momentum $\theta_0$
(see Appendix~\ref{sec:twist}).
Thus, at $t=0$, $n_q$ is a quasi (broadened) Fermi distribution
with the shifted Fermi wave vectors $\pm k_F' = \pm\pi/2+\theta_0$.
After the flux is quenched to zero, the distribution starts evolving.

\begin{figure*}
   \includegraphics[width=8cm]{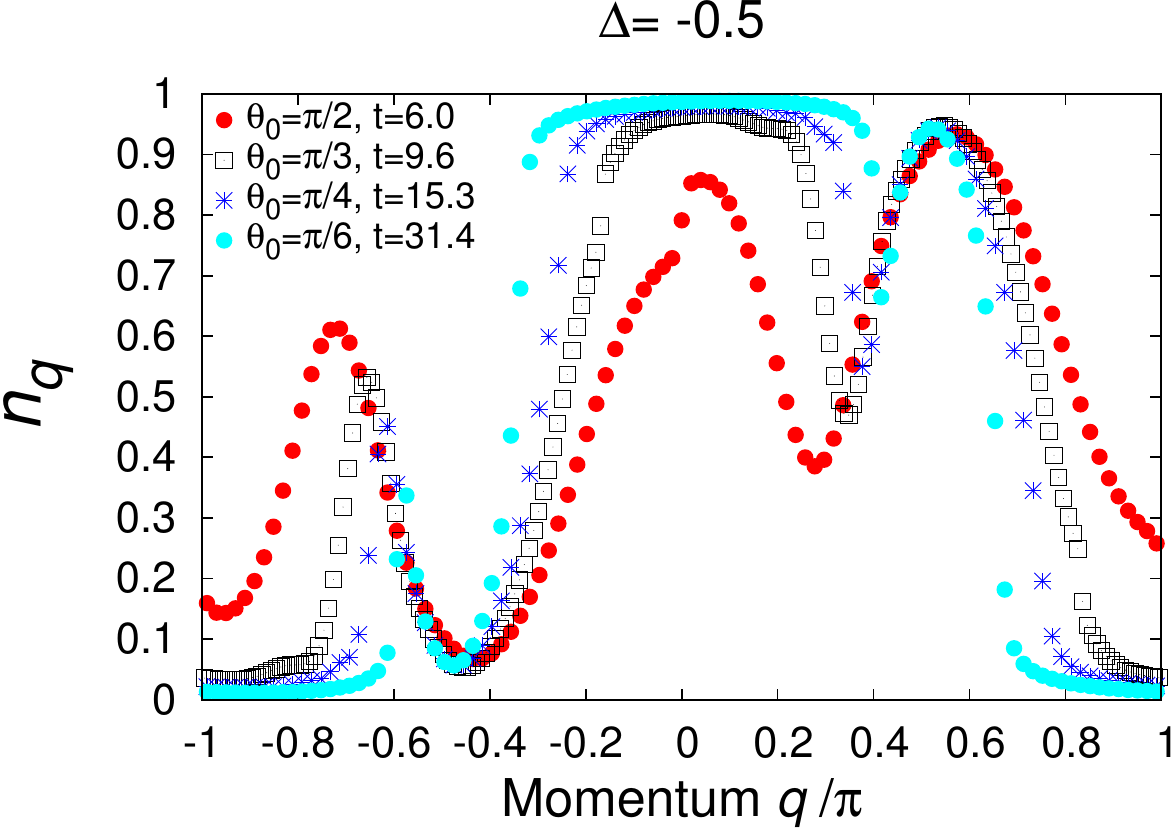}
    \includegraphics[width=8cm]{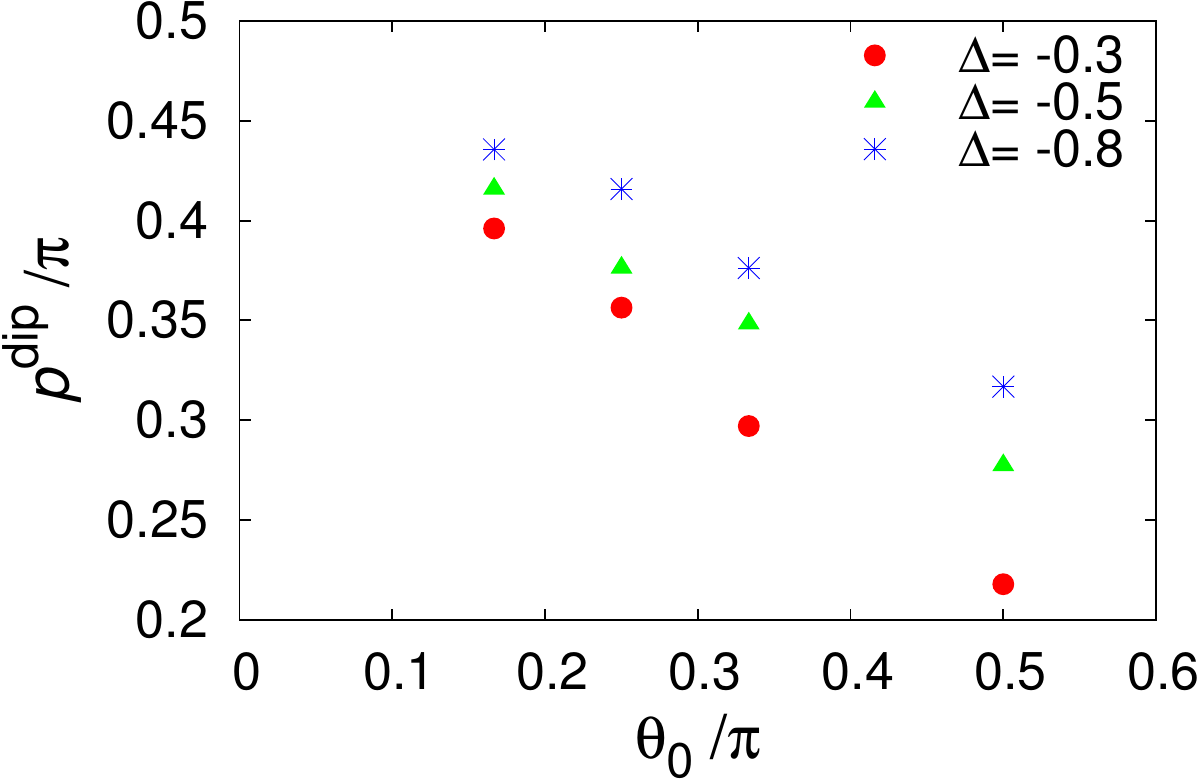}
  \caption{   \label{Dip_comp}
  Left: Momentum distribution for $\Delta=-0.5$ with different $\theta_0$.
  The time is chosen to match the first minimum of the current after the quench.
  Right: Momentum of dip, $p^{\rm dip}$, versus $\theta_0$.
    }
\end{figure*}

An important observation  is the simultaneous appearance of a  ``dip''  and a ``peak'' 
in $n_q$ (left of Figs.~\ref{TimeEvol_MomD} and~\ref{Dip_comp}). Both structures 
appear to oscillate in phase with $J(t)$, as shown in the inset of Fig.~\ref{TimeEvol_MomD}.
We also note that, in situations
where current oscillations are absent (right panel of Fig.~\ref{TimeEvol_MomD}), no dip/peak is observed.
The dip and the peak correspond to two momenta $p^{\rm dip}$ and $p^{\rm peak}$
that are separated by $\pi$: $p^{\rm peak} = p^{\rm dip} -\pi$.
The  momentum $p^{\rm dip}$ is plotted as a function of $\theta_0$ for
several values of $\Delta$ in the right of Fig.~\ref{Dip_comp}.

For small $\theta_0$, $p^{\rm dip}$ approaches the shifted Fermi point $k_F'$ ($\to\pi/2$) and the
current-carrying modes become low-energy modes. This is expected since the initial state and the ground
state are energetically close to each other in this case.
However for finite $\theta_0$ we observe that the modes responsible for the current oscillations
are located at a significant distance from the Fermi points, and are not low-energy excitations.

The detailed dependence of $p^{\rm dip}$ on the  parameters $\theta_0$ and $\Delta$ is not yet understood
but we can consider a simplified picture where only two characteristic modes
govern the current dynamics. Located at $p^{\rm dip}$
and $p^{\rm peak}=p^{\rm dip}-\pi$, these modes are related through some Umklapp processes induced by the interactions.

As mentioned above, since $p^{\rm dip}$ generically departs from $k_F'$,
the oscillations might not be described by the Tomonaga-Luttinger liquid
(TLL) framework, where the physics is entirely described in terms of
low-energy excitations in the vicinity of the Fermi points
\cite{GiamarchiBook}.  As a comparison, a different global quench for
the XXZ model (equivalent to the present Hamiltonian) was considered in
Refs.~\onlinecite{Pollmann2013} and~\onlinecite{Karrasch2012}.  There,
the strength of the interaction is suddenly changed and several aspects
of the dynamics appeared to be well described by the TLL model.

Finally, we point out that this anomalous dip (peak) structure might be observed in real experiments,
since the momentum distribution  is often accessible in cold atom experiments.

To summarize this subsection, we  found numerically that 
the oscillation frequency is proportional to the strength of interaction $|\Delta|$ and that
these oscillations are governed by excitations located in momentum space far from the shifted Fermi point.

\subsection{The long-time limit}
\begin{figure*}
   \includegraphics[width=8cm]{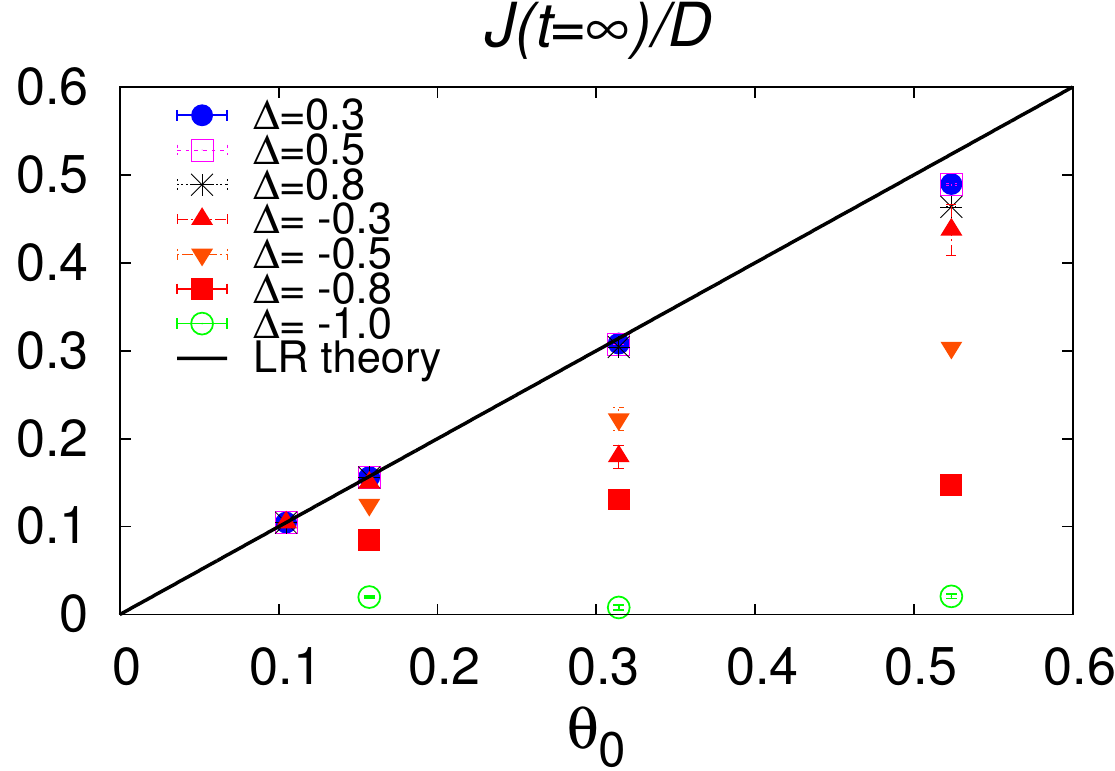}
   \includegraphics[width=8cm]{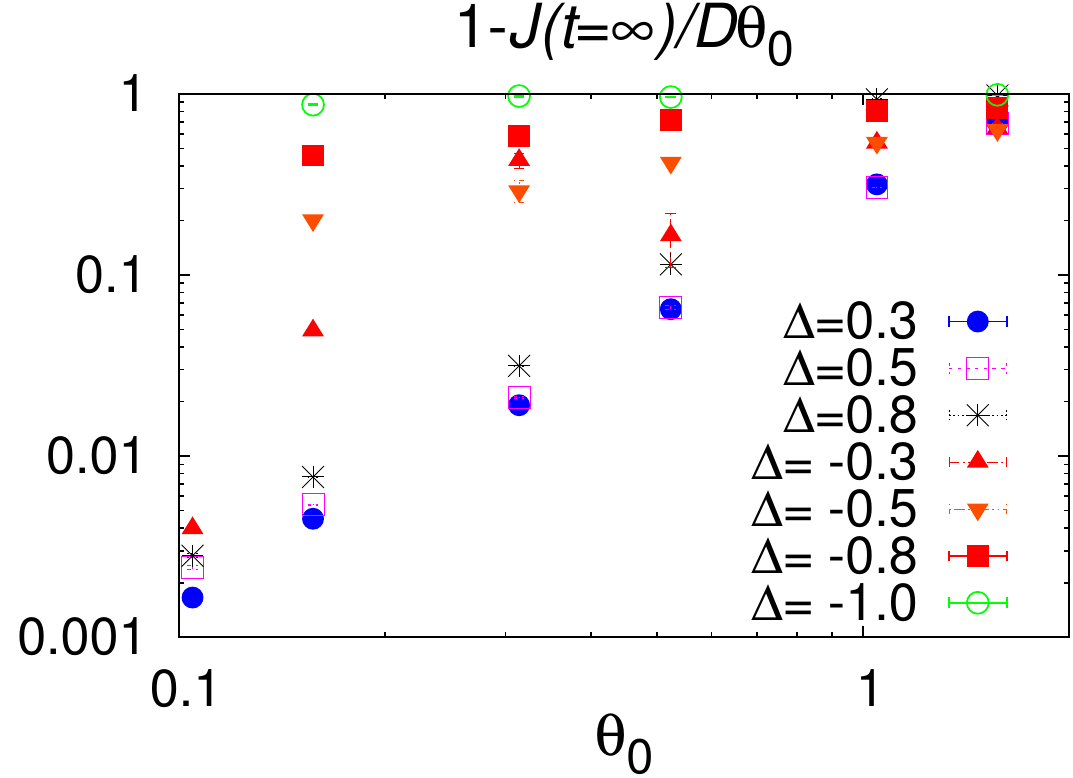}
   \caption{ \label{J_infinity} Numerical data for (left) $J(t=\infty)/D$  and
                 (right) normalized deviation from LR theory $1-J(t=\infty)/(D\theta_0)$.}
\end{figure*}
Here we discuss the long-time limit of the current.
From the LR theory in $\theta_0$,  the long-time limit of the current $J(t=\infty)$ is
given by $J(t=\infty) = D \theta_0$, where $D$ is the Drude weight of the system.
The latter is exactly known for the XXZ model (equivalent to our present model)
at zero temperature~\cite{Shastry1990}:
\begin{equation}
 D = \frac{\pi}{4} \, \frac{\sin\mu}{\mu(\pi-\mu)},
\:\:\: \mu = \arccos(-\Delta).
\end{equation}
$D$ is non-zero only in the gapless phase ($-1 \leq \Delta < 1$), and
vanishes in the gapped phase. Hence the LR theory predicts that
$J(t=\infty)$ is non-zero only in gapless phase. In addition, $D$ is
symmetric under $\Delta \leftrightarrow -\Delta$ (except for $\Delta =
\pm 1$).  So the long-time limit of the current does not depend on the
sign of $\Delta$ in the LR theory.

We can also estimate the long-time limit of
the current, $J(t=\infty)$, by fitting and extrapolating the numerical data
obtained for the finite time after the quench by the iTEBD method.
In the following, we shall compare $J(t=\infty)$ estimated from
the iTEBD calculation and that predicted by the LR theory.
Note however that  in the case of $\Delta<0$ and small initial flux,
it is difficult to evaluate the long-time limit from numerical data.

As shown in Fig.~\ref{Result1}, the numerical results indicate that 
$J(t=\infty)$ is non-zero in the gapless phase and  zero in the gapped phase,
which is consistent with the LR result.
On the other hand, it is clear from Fig.~\ref{J_infinity} that
$J(t=\infty)$ rapidly deviates from the LR prediction when $\theta_0$ increases.
This reflects the nonlinearity of current as a function of
the initial flux strength.  For attractive interactions ($\Delta>0$),
the normalized deviation from the LR theory (right of
Fig.~\ref{J_infinity}) shows power-law decay at small $\theta_0$, and
these appear to be compatible with
$J(t=\infty)=D\theta_0+\mathcal{O}(\theta_0^3)$.
In contrast, for large repulsive interactions ($\Delta<0$),
$J(t=\infty)$ strongly deviates from the LR theory even for $\theta_0$
as small as $\pi/30 \approx 0.1$.
We note that, despite the general difficulty in doing such fits and
extrapolations,
the deviation from the LR theory cannot be attributed to some
error in the extrapolation.
This is clear by comparing the raw finite-time data and the LR
theory prediction, as shown in Fig.~\ref{ExampleOfFit}; any sensible extrapolation
would give different $J(t=\infty)$ from the LR theory.
We also note that this strong non-linearity in presence of repulsive interactions
has already been noted in Ref.~\onlinecite{Peotta2014}.

It is an interesting fact that the magnitude of the above nonlinearities strongly depends on the sign of $\Delta$.
We may attribute this to 
{\it superfluid} correlations in the system.
Strictly speaking, superfluidity is absent in one dimension, but a superfluid-like response can be observed
in some  dynamical properties of the system~\cite{ThomasOshikawa2011},
and these are expected to be stronger for attractive interactions ($\Delta>0$) than for repulsive ones ($\Delta<0$).
In the case of repulsive interactions,
the larger normal component dissipates and this would result in smaller values
of $J(t=\infty)$, as observed in Fig.~\ref{J_infinity}.

\subsubsection{Comparing the momentum distributions
 in the gapless phase and in the gapped phase}

\begin{figure*}
   \includegraphics[width=8cm]{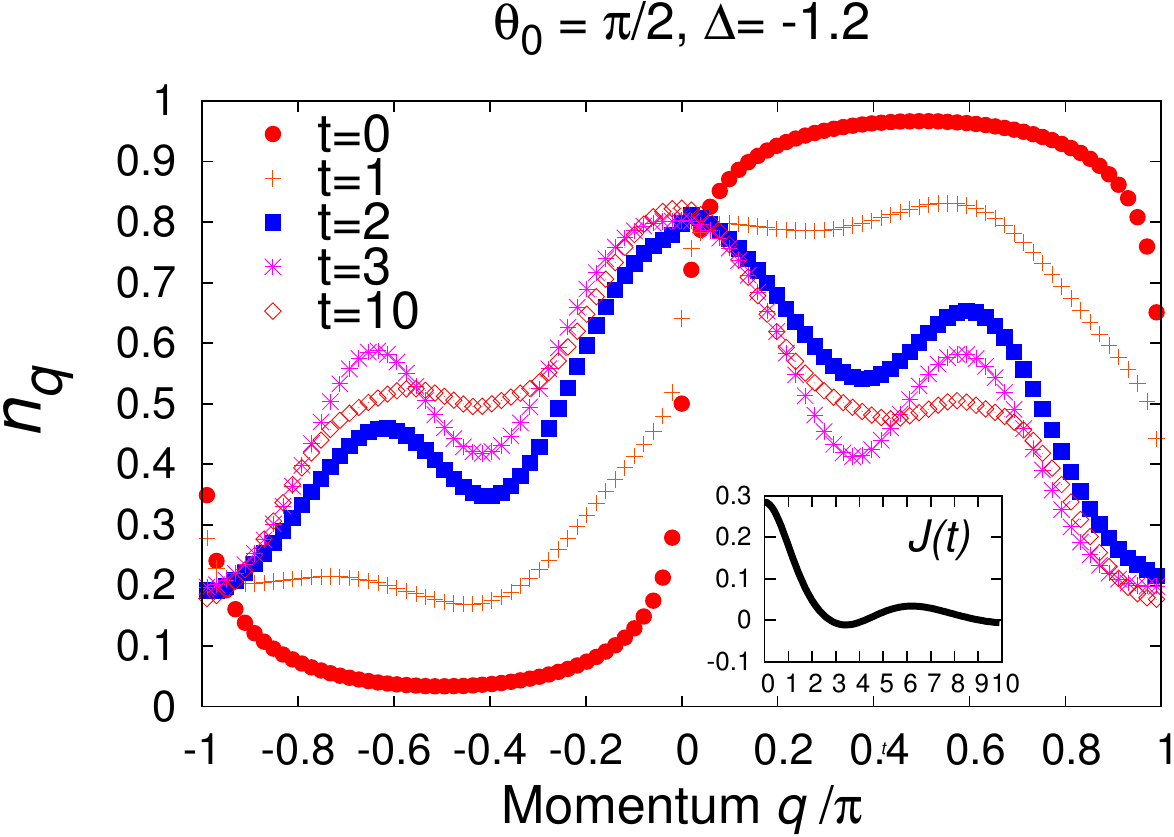}
    \includegraphics[width=8cm]{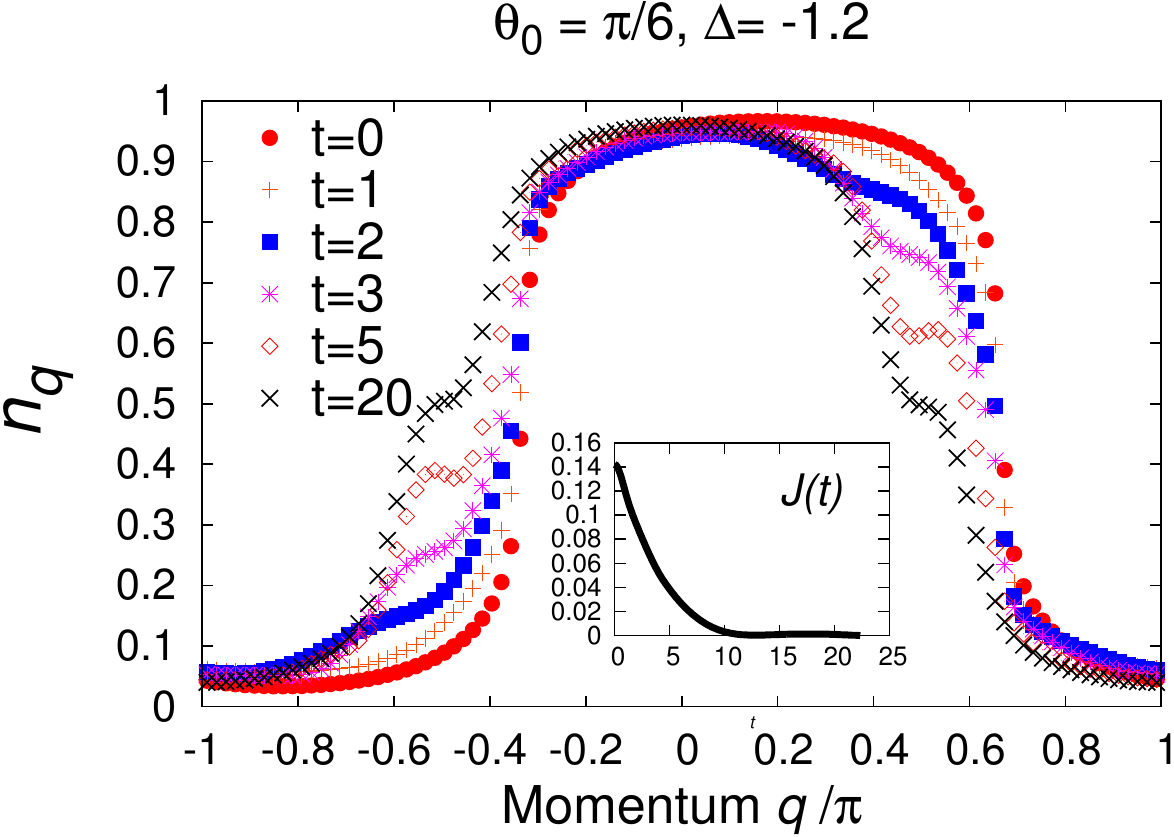}
  \caption{  \label{TimeEvol_MomD_gapped}
  Time evolution of the momentum distribution in the gapped phase.
  Left panel: $\theta_0= \pi/2,\,\Delta=-1.2$.  Right panel: $\theta_0= \pi/6,\,\Delta=-1.2$.
  In both cases, the whole momentum distribution moves towards the center ($q=0$)
  and the current decays to zero.}
\end{figure*}

The dynamics of the momentum distribution shows a qualitative difference between the  gapless and the gapped phases, irrespective of initial flux $\theta_0$.
In the gapless phase, the shifted Fermi sea structure of the
initial state appears to be qualitatively robust and survives
up to the stationary regime
(Fig.~\ref{TimeEvol_MomD}).
The imbalance between the number of left ($q>0$) and right ($q<0$)
moving particles in the stationary regime is the source of the persistent current $J(t=\infty)\neq 0$. 
On the other hand, in the gapped phase, the shifted Fermi sea structure of the  initial state
disappears over a relatively short time scale (Fig.~\ref{TimeEvol_MomD_gapped}).
In that case, the whole momentum distribution moves towards the center ($q=0$)
and the symmetry between $q>0$ and $q<0$ is restored, leading to $J(t=\infty)=0$.

This  difference between the gapless and gapped phases
might be related to the presence of additional
conserved quantities that exist in the gapless phase~\cite{ProsenPRL2011,ProsenPRL2013,pereira_exactly_2014,ProsenNPB2014}.
Those additional constants of motion, called quasi-local, are responsible for the ballistic transport and the 
non-zero Drude weight at finite temperature in the gapless phase~\cite{ProsenPRL2011}.
In a similar manner, we expect these additional conserved quantities to prevent the restoration of the left-right symmetry in the momentum distribution.

\subsection{Relaxation time}

\begin{figure*}
   \includegraphics[width=8cm]{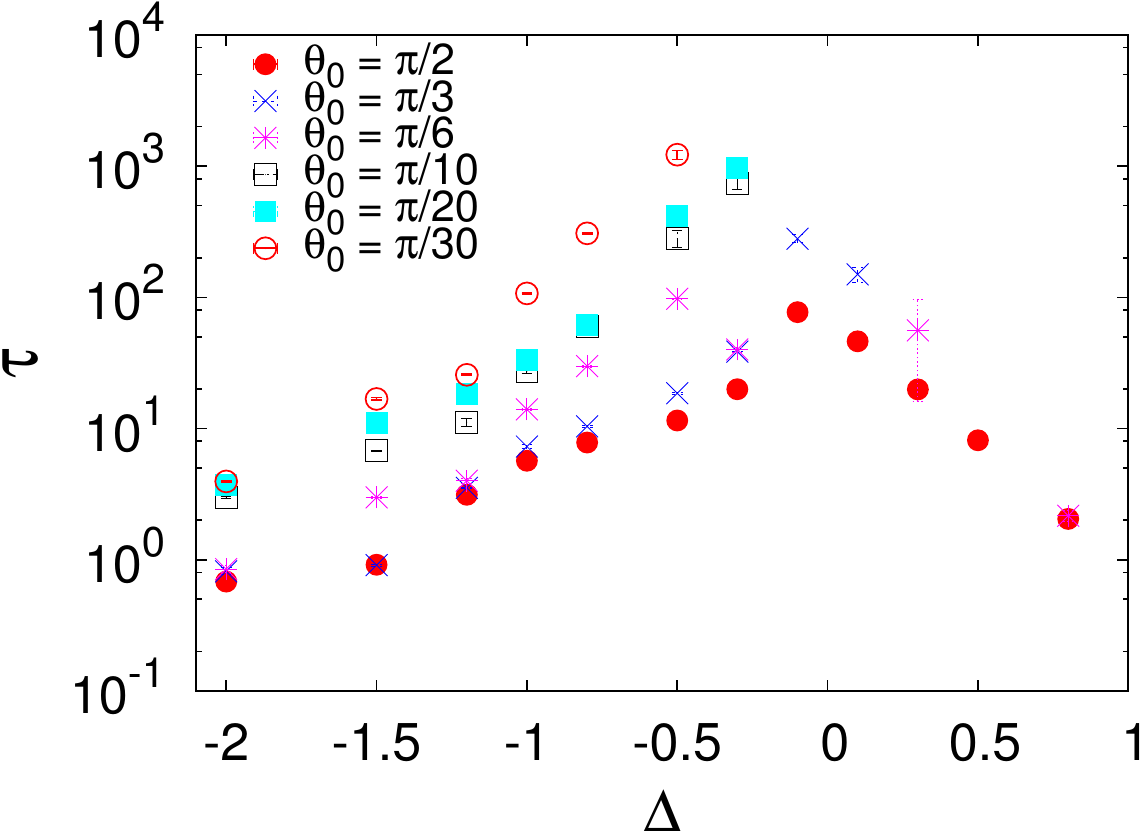}
  \includegraphics[width=8cm]{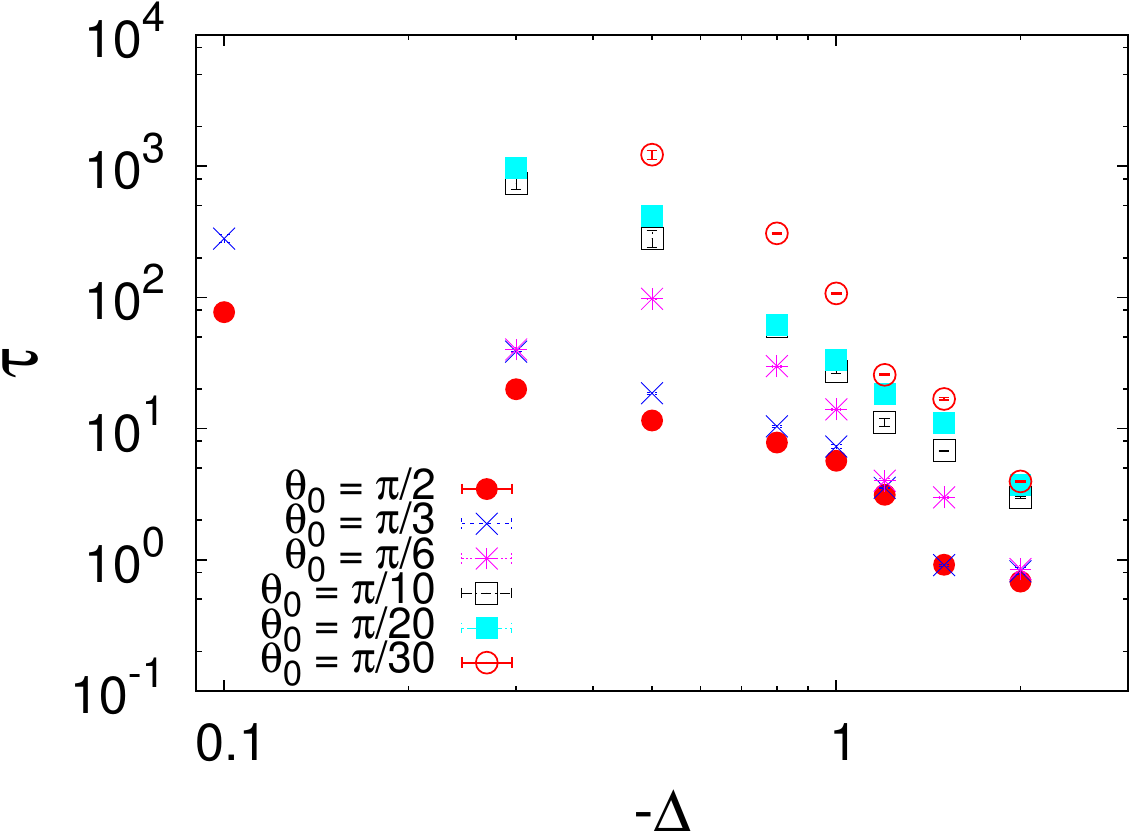}
  \caption{\label{Tau} Relaxation time $\tau$ versus the interaction parameter $\Delta$.
     Left panel: semi-log plot. Right panel: log-log plot.
  The data points where the decay is too slow to be reliably quantified with our calculations are omitted.}
\end{figure*}

The relaxation time $\tau$, extracted from the fits, is plotted in Fig.~\ref{Tau}.
In general,
larger $|\Delta|$ and larger $\theta_0$ result in smaller $\tau$ (faster decay).
This is natural because the time derivative of
the current $dJ(t)/dt$ is proportional to $\Delta$ (Eqn.~(\ref{dJdt})).
When $\Delta=0$ (free fermion point), the current is conserved and
$\tau$ must be $\infty$. In agreement with this fact, $\tau$ appears to diverge when
$|\Delta| \to 0$.

\section{Conclusion}\label{Conclusion}
We have studied numerically a flux quench in an interacting spinless fermion model in one dimension.
This quench generates some particle current at the initial time and we monitored and analyzed quantitatively the current dynamics that follows.
The numerical data reveal some current oscillations as well as some decay to a stationary value.
For  repulsive interactions and a large initial flux, the frequency of those oscillations is proportional
to the strength of the interaction in the system. Remarkably, the dynamics of the momentum distribution reveals that 
these oscillations are governed by excitations located deep inside the (shifted) Fermi sea.
In addition to those novel oscillations, the long-time limit of the current
exhibits nonlinearities which are particularly strong in presence of repulsive interactions.

As future work, it seems important to understand the origin of the specific ``dip'' momentum $p^{\rm dip}$
that governs the current oscillations. Another interesting direction of research would be to
compute  the long-time limit of the current (beyond the weak-flux regime where the linear response theory applies)
using integrability techniques~\cite{de_luca_quenching_2014,ProsenNPB2014,pereira_exactly_2014}.

\begin{acknowledgments}
 YON  acknowledges Y. Tada, I. Danshita, S. Furukawa, and S. C. Furuya for
 valuable comments and discussions, and is supported 
 by Advanced Leading Graduate Course for Photon Science (ALPS) of JSPS. 
 YON and MO thank the Yukawa Institute for Theoretical Physics at Kyoto University (YITP).
Discussions during the YITP workshop YITP-W-14-02 on
``Higgs Modes in Condensed Mater and Quantum gases'' were useful to complete this work.
MO and YON are supported by MEXT/JSPS KAKENHI Grant Nos. 25103706 and 25400392.
 GM acknowledges V. Pasquier for useful discussions, and is supported by
 a JCJC grant of the Agence Nationale pour la Recherche (Project
 No. ANR-12-JS04-0010-01).
The computation in this paper has been partially carried out by using the
facilities of the Supercomputer Center, the Institute for Solid State
Physics, the University of Tokyo. 
\end{acknowledgments}

\appendix

\section{Initial state and twist operator}
\label{sec:twist}
As mentioned in the introduction, in a ring of length $N$, the zero-flux Hamiltonian $H_0$ and the Hamiltonian with $p\in\mathbb{Z}$ flux quanta
are related by a unitary transformation:
\begin{equation}
 H_{\theta_0=2\pi p/N} =  U_p H_0  U_p^{-1},
\end{equation}
where $U_p$ (so-called twist operator) is defined as \cite{lieb_two_1961}
\begin{equation}
 U_p=\exp\left(\frac{2ip \pi}{N}\sum_{x=0}^{N-1} x c^\dagger_x c_x \right).
\end{equation}

So, the ground state $\left|\psi_p \right\rangle$ in presence of $p$ flux quanta can be expressed in terms of the 
zero-flux ground-state $\left|\psi_0 \right\rangle$:
\begin{equation}
\left|\psi_p \right\rangle=U_p \left|\psi_0 \right\rangle.
\end{equation}
In other words, when $p$ is an integer, the flux quench amounts to study the dynamics generated by $H_0$ when starting from the initial
state $U_p \left|\psi_0 \right\rangle$.

$U_p$ has a simple action on the fermion creation operators:
\begin{equation}
 U_p c^\dagger_x U_p^{-1} = \exp\left(\frac{2ipx \pi}{N}\right)   c^\dagger_x,
\end{equation}
which implies that it performs a momentum shift (or boost):
\begin{equation}
  U_p \tilde c^\dagger_q U_p^{-1} =  \tilde c^\dagger_{q+2\pi p/N}.
\end{equation}
In the case of a noninteracting fermion problem ($\Delta=0$
in Eqn.~\eqref{DefOfModel}), $U_p$ maps the Fermi sea $\left|\psi_0 \right\rangle$
to a ``shifted Fermi'' sea $U_p \left|\psi_0 \right\rangle$,
with Fermi points located at $-\pi/2+2\pi p/N$ and $+\pi/2+2\pi p/N$. The latter is an exact excited eigenstate of
the Hamiltonian $H_0$. Therefore, in the noninteracting case, the flux quench does not
generate any dynamics.

\section{Numerical calculations}
\label{sec:app_num}
This appendix provides some  details on the numerical method.
We first prepare the ground state of 
the Hamiltonian with flux $\theta_0$
\begin{equation} 
 H_{\theta_0} = - \frac{1}{2} \sum_i \left( 
e^{-i\theta_0} c^\dagger_i c_{i+1}  + \text{h.c.}  \right)
- \Delta \sum_i  \tilde{n}_i \tilde{n}_{i+1},
\end{equation}
using an imaginary time-evolution (with iTEBD). A second-order Suzuki-Trotter decomposition is used
and we take the bond dimension $\chi$ between 500 and 1200.
The imaginary time step $\delta \tau$ is reduced gradually from $\delta \tau = 0.1$ to $\delta \tau = 0.001$.
$\delta\tau$ is reduced each time the imaginary time propagation with a coarser $\delta\tau$ has converged.
At each $\delta\tau$, the convergence is checked by looking at the energy
and the entanglement entropy of a half chain between two successive time steps.
Our convergence criterion is $10^{-8}$ for the energy and $10^{-6}$ for the entanglement entropy.
After the imaginary time-evolution at $\delta\tau=0.001$
converges, we compare the obtained energy with the exact one~\cite{TakahashiBook}.
Our  iTEBD energy matches the exact value with 5 or 6 digits.
Note  that the entanglement entropy of a half-infinite system should diverge in the gapless phase,
and that it is approximated here by a finite value (since $\chi$ is finite).
See Refs.~\onlinecite{tagliacozzo_scaling_2008} and  \onlinecite{Pollmann2013} for related discussions. 

\begin{figure*}
   \includegraphics[width=8cm]{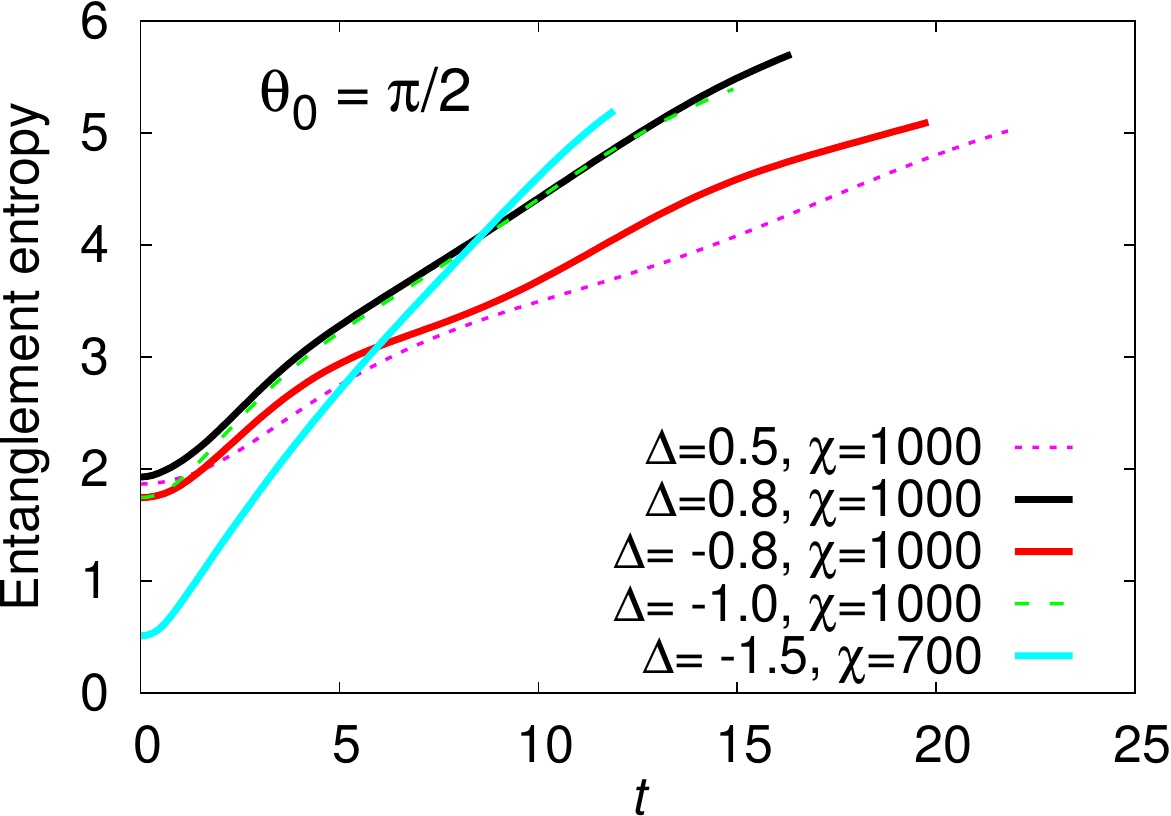}
   \includegraphics[width=8cm]{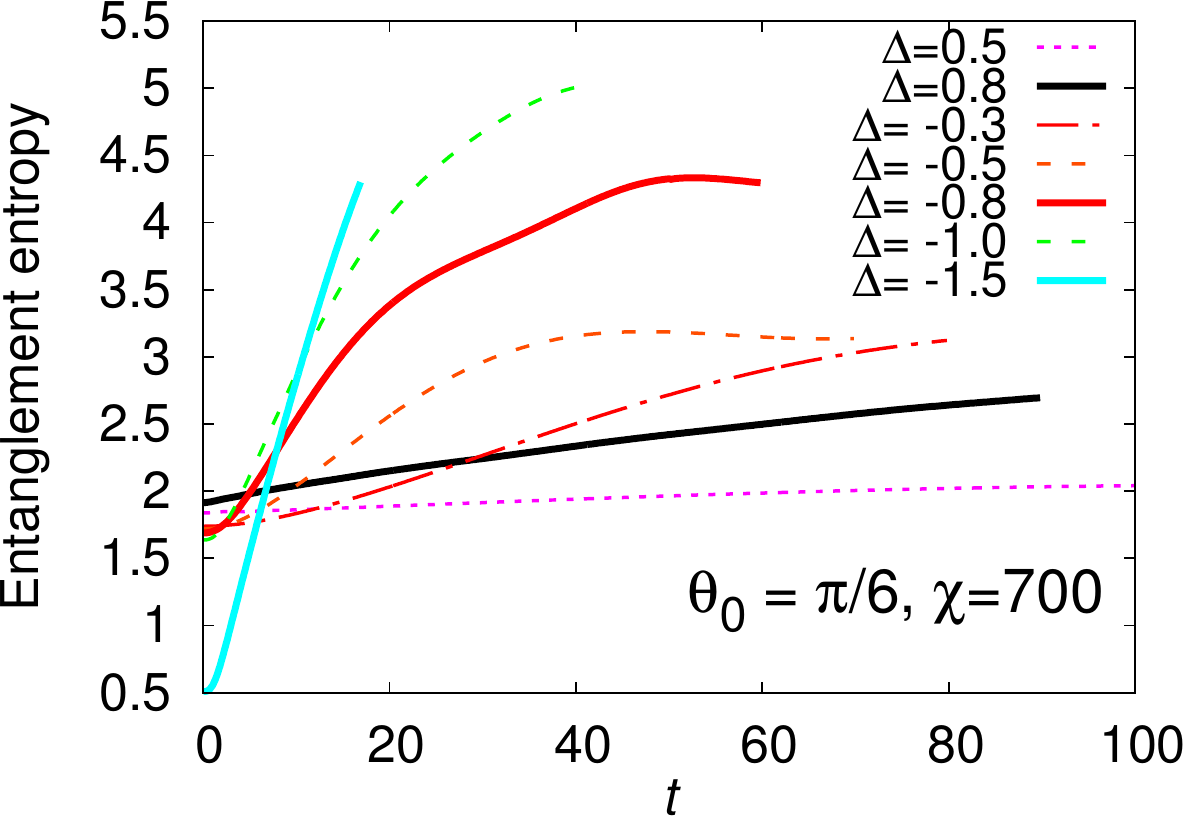}
   \caption{   \label{Growth_EE}
   Growth of entanglement entropy of a half chain for $\theta_0=\pi/2$ (Left) and $\theta_0=\pi/6$ (Right) and various values of $\Delta$. 
   One can check that the (weak) oscillations of the entanglement entropy are in phase with the oscillations of the current.}
\end{figure*}

Next, using iTEBD again, we calculate the real time-evolution using the Hamiltonian without flux:
\begin{equation} 
 H_0 = - \frac{1}{2} \sum_i \left( 
 c^\dagger_i c_{i+1}  + \text{h.c.}  \right)
- \Delta \sum_i  \tilde{n}_i \tilde{n}_{i+1}.
\end{equation}
We again use a second-order Suzuki-Trotter decomposition
and take a real-time step  $dt=0.01$ or 0.02.
One of the largest obstacles to calculate the real time-evolution of a quantum system by iTEBD
is the  growth of the entanglement entropy. This growth is usually linear in time for global quenches,
and, as shown in Fig.~\ref{Growth_EE}, it  appears to be the case for the present flux quench.
In practice this has limited the accessible time scale to $t\lesssim 100$.

\begin{figure}
   \includegraphics[width=8cm]{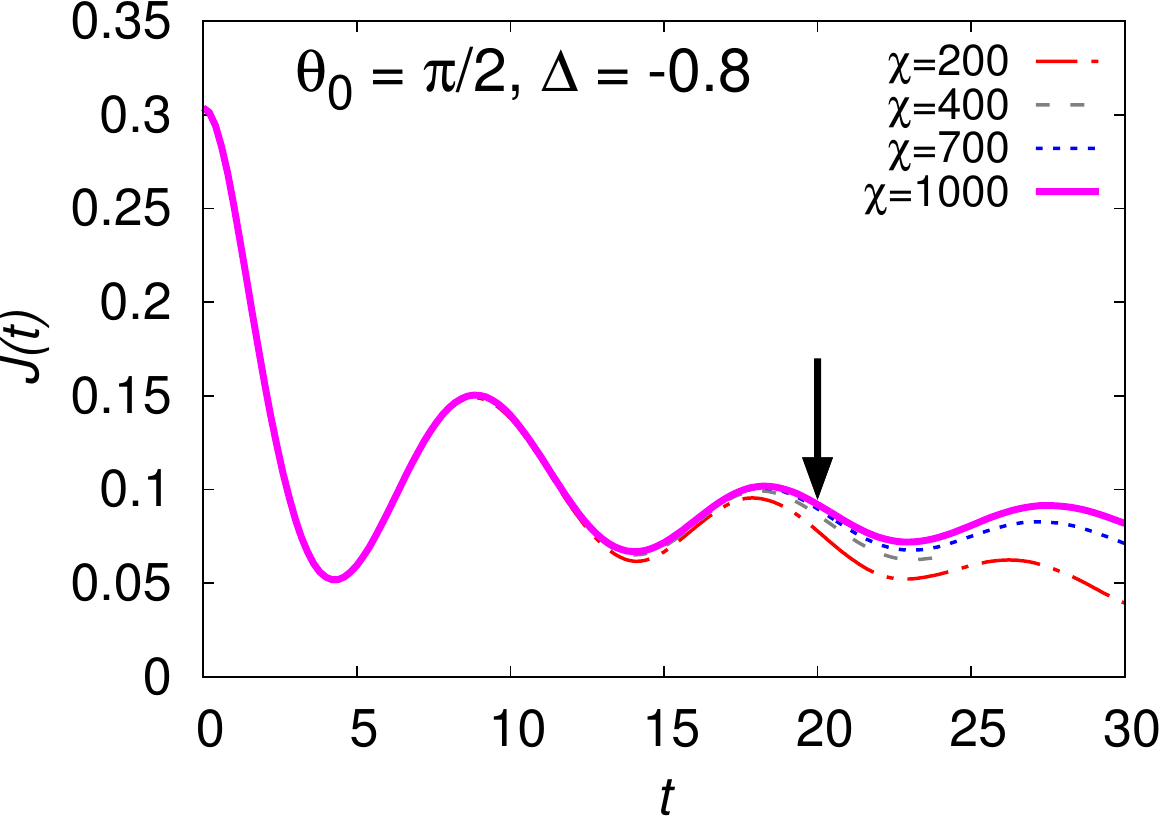}
   \caption{\label{dynamics_vs_chi}
   Expectation value of the current, computed with several values of the bond dimension $\chi$. In this case
   we  retain the data for $\chi=1000$, and only up to $t=20$ (indicated by arrow).}
\end{figure}

In order to check the accuracy of the simulated time-evolution,
we have monitored the accumulated truncation errors, the
conservation of energy, and the 
conservation of  the number of particles.
In addition to monitoring these values, we confirmed the accuracy of our results for a few $\Delta$ and $\theta_0$
by comparing the results for several values of the time step $dt$ and the bond dimension $\chi$. 
As an example, Fig.~\ref{dynamics_vs_chi} shows how the expectation value of the current is modified
when increasing the bond dimension.

\bibliography{Manuscript_v3_arXiv}

\end{document}